\documentstyle[12pt,preprint,aps]{revtex}

\tightenlines
 
\begin{document}
 
\draft
 
\title{Multifractality in Time Series}
\author{Enrique Canessa\thanks{E-mail: canessae@ictp.trieste.it}}
\address{The Abdus Salam ICTP, International Centre for Theoretical
Physics, \\
P.O. Box 586, 34100 Trieste, Italy}
\date{\today}
 
\maketitle
\begin{abstract}
\baselineskip=20pt

We apply the concepts of multifractal physics to financial
time series in order to characterize the onset of crash for the
Standard \& Poor's 500 stock index $x(t)$.  
It is found that within the framework of multifractality,
the "analogous" specific heat of the S\&P500 discrete price index displays
a shoulder to the right of the main peak for low values of time lags.
On decreasing $T$, the presence of the shoulder is a consequence
of the peaked, temporal $x(t+T)-x(t)$ fluctuations in this regime.
For large time lags ($T > 80$), we have found that $C_{q}$ displays typical
features of a classical phase transition at a critical point.
An example of such dynamic 
phase transition in a simple economic model system, based on
a mapping with multifractality phenomena in random multiplicative
processes, is also presented by applying former results obtained with a 
continuous probability theory for describing scaling measures.

Pacs: 64.60.Ak, 02.50.+s, 89.90.+n, 01.75.+m

\vspace{2.0cm}
\hfill{ To appear in \em{J. Phys. A: Mathematical and General (2000)}}
\end{abstract}

\newpage

\baselineskip=22pt

\section{introduction}

In this work we apply the concepts of multifractal physics to financial
time series in order to characterize the onset of crash for the
Standard \& Poor's 500 stock index.  We shall present an example of
dynamic phase transition in a simple economic model system based on
a mapping with multifractality phenomena in random multiplicative
processes and by applying former results obtained with a continuous
probability theory for describing scaling measures.
 
An attempt is made to characterize the presence of stock market
crashes by solving for a price equation from a non-linear equilibrium
model and showing how multifractal physics measures can be generated
from this equation.  We found that an "analogous" specific heat
$C_{q}$ of the S\&P500 price data displays a shoulder to the right
of the main peak as a function of time lags.  For large time lags,
$C_{q}$ resembles a classical phase transition at a critical point.
We explain this dynamic phase transition by a mapping with multifractality
phenomena in random multiplicative processes. Within this description
the temporal price variations of a commodity displays
features of an "analogous" phase transition from inflated to devalued
prices, when the excess demand is not linear in the asset price.  An
analytical expression for $C_{q}$ of the economic model system is
derived.

Finance and physics joint "ventures" have attracted considerable
interest in the literature for over many years \cite{Cha98}.
These efforts have allowed to pursue analogies between stock
market dynamics and stochastic models commonly used in the statistical
physics of complex systems \cite{And88}.  Such parallel analysis have
been useful to best quantify and understand possible correlations in
financial data by measuring the autocorrelation function
\cite{Con97} and the power spectrum \cite{Liu97}.  Another notable example
is the analogy with the scaling properties observed in turbulence
\cite{Gha96,Man96,Arn98}.

Parallels have also been drawn between a very simple theory of financial
markets and the quantum gauge theory \cite{Ili97} (see also comments in
\cite{Sorn98}).  Other interesting
studies for investment strategies in diversified portfolio stocks have
been recently discussed in terms of products of random matrices \cite{Gal97}
and of multiplicative random walks \cite{Mar98}.  Realistic price
fluctuations have also been found to emerge in the adaptation of a system
({\em i.e.}, the traders) to complex environments \cite{Pot96} and from the
self-organization of a (closed) system of traders without external
influences \cite{Cal97}.

Thus, the financial market dynamics is still an open subject for physicists
and economists.  In particular, the basic principles governing the origin
of stock market crashes are far from well understood by both communities
\cite{Sor97}, specially in regard to the problem of world-wide market
dynamics.  Using the renormalization group theory
\cite{Sor97,Fei95,Sor96,Glu97}, it has been proposed that this cooperative
phenomenon could be the result of a critical phase transition
(see also \cite{Van98}).  Motivated
by these suggestions, it is then tempting to determine the constrains
needed to understand and describe analytically an analogous phase transition
from a new perspective.  That is, by using the characterization of 
multifractal singularities where an analogy with thermodynamics has
already been established \cite{Hal86,Mea98}.

Multifractality was initially proposed to treat turbulence and, in
recent years, it has been applied successfully in many different fields
ranging from model systems such as Diffusion Limited Aggregation to
physiological data such as heartbeat.  The multifractal analysis
of financial discrete sequences developed in our paper is another aspect in
a relatively new topic of physics, the so-called econophysics.

\section{Multifractal Characterization}

This paper will focus on scaling laws in financial returns, particularly on the
absolute values of returns and their scatter as a function of time.   
Similarly to other studies \cite{Man99},
we are not attempting to predict a price drop or rise on a specific time
on the basis of past records,
but we are after a new characterization of the presence of stock crashes, 
which is the added value of our distinctive approach.
This is important to mention since forecasting has been extensively
studied in the econometrics literature by other
techniques such as ARCH and multivariance models depending, {\it e.g.}, on
the data seasonal behavior and previous experiences of the forecasters.
Interesting work on fractional processes of these types in econometrics 
is very well known and these are beyond the scope of our work.  
Important to mention is that recent theoretical work
has considered the possibility of fractal nature of the absolute value 
of returns on exchange rates \cite{Man99}.

Let us introduce next our new characterization of a financial time series
$x(t)$ based on the well-known definitions used in multifractal
physics.  In Fig.1 we show the temporal $x(t+T)-x(t)$ fluctuations
of the S\&P500 index (data available in \cite{Ley96}) for the 12 years
period 2 January 1980 - 31 December 1992 as a function of the
trading time lags $1 \le T \le 220$.  From this figure, it can be
observed that the peaked narrow fluctuations measured for $T=1$
spread out on increasing $T$.  For $T \rightarrow 80$, the maximum and
minimun difference values of the so-called {\it "Black Monday"} crash
measured in 1987 (shown as the largest straight line for $T=1$ in the
figure), become comparable to the valleys and picks differences of the
relative S\&P500 fluctuations over wide periods of time.  This peculiar
behavior is shown next to lead to an analogous thermodynamic phase
transition when varying $T$.

Let us consider the following measure over $N$ intervals
\begin{equation}\label{eq:mui}
\mu_{i}(T) = \frac{|x(t+T)-x(t)|}{ \sum_{t=1}^{N}|x(t+T)-x(t)|}
  \;\;\; ,
\end{equation}
with $T$ representing some finite trading time lags $T$.  Clearly,
the above relation can be viewed as a normalized probability measure
with $\mu_{i} > 0$.

From $\mu_{i}$ in Eq.(\ref{eq:mui}), we then construct the corresponding
generating function $Z$, and its moments $q$, which follows the scaling
\begin{equation}\label{eq:zqn}
Z(q,N)=\lim_{T\rightarrow \infty} \sum_{t=1}^{N} \mu_{i}(T)^{q}
    \sim N^{-\tau(q)}\;\;\; .
\end{equation}
To get a thermodynamic interpretation of multifractality (see,
{\it e.g.}, \cite{Hal86,Mea98,Ber98}), we divide
the one-dimensional system of length $L$ into $N$ lines of
length $\ell$; thus $N \sim L/\ell$.  We then associated this 
$N$ with the number of discrete $x(t)$ time sequences considered in 
Eq.(\ref{eq:mui}) in order to relate $T$, $L$ and $\ell$ in the
definition of the measure.  For $\ell/L \rightarrow 0$,
the function $\tau$ relates to the generalized fractal dimensions
$(q-1) D_{q}$.  Similar multifractal analysis
has recently been performed for the energy dissipation field of
turbulence \cite{Men87}, logistic maps \cite{Kur97} and surface
roughening \cite{Ber98}.

By following the thermodynamic formulation of multifractal measures,
we can also derive an expression for the "analogous" specific heat
as follows
\begin{equation}\label{eq:cqm}
C_{q}  \equiv -\; \frac{\partial^{2}\tau (q)}{\partial q^{2}}
  \approx  \tau(q+1) - 2 \tau(q) + \tau(q-1) \;\;\; .
\end{equation}
For large $T$, we shall show that the form of $C_{q}$ resembles a 
classical phase transition at a critical point.

We shall return to these equations in Section III.

\section{Dynamical Model}

Similarly to \cite{Cal97,Bou98} in our approach there is only one stock.
To study the price changes for one commodity it is necessary
to derive a dynamical equation which results from the prevailing market
conditions.  The market is usually considered competitive so it self-organizes
to determine the behaviour of prices.  We assume here that all factors
determining the demand $D$ and the supply $Q$ other than the asset price
$p$ remain constant over time and denote these quantities in equilibrium
with an asterisks ($^{*}$).  In the following all variables are dimensionless.

In a competitive market it is expected that the rate of price increase
should be a functional of the excess demand function $E(p)=D(p)-Q(p)$.
Hence one writes $dp/dt \equiv f[ E(p) ]$ \cite{Cle84}.  Assuming that
a commodity can be stored then, in general, the flow of demand does not
equal the flow of $Q$ output.  Hence stocks of the commodity (or product)
build up when the flow of output exceeds the flow of demand and vice-versa.
Then the rate at which {\it the level of stocks} $S$ changes 
can be approximated as $dS/dt = Q(p) - D(p)$.  From these relations, a price
adjustment relation that takes into account deviations of the stock
level $S$ above certain optimal level $S_{o}$ (to meet any demand reasonably
quickly) is then given by
\begin{equation}\label{eq:stock}
\frac{dp}{dt} = -\gamma \frac{dS}{dt} + \lambda (S_{o}-S)  \;\;\; ,
\end{equation}
where $\gamma$ ({\it i.e.}, the inverse of excess demand required
to move prices one unity \cite{Bou98}) and $\lambda$ are positive
parameters.  If $\lambda = 0$, the price adjusts at a rate proportional
to the rate at which stocks are either raising or running down.  If
$\lambda > 0$, prices would increase when stock levels are low and
rise when they are high (with respect to $S_{o}$).  We shall assume
here $\lambda$ to characterize a noise term.

In our description, for each asset price $p$, we postulate simple
non-linear forms for the quantities $D$ demanded and $Q$ supplied such
that
\begin{eqnarray}\label{eq:dq}
D(p) & = &  d^{*} + d_{o} [ \; 1 - \frac{\delta^{2}}{2!}(p-p^{*})^{2} + 
       \dots \;] (p-p^{*}) \;\;\; ,
\nonumber \\
Q(p) & = &  q^{*} + q_{o} [ \; 1 - \frac{\delta^{2}}{2!}(p-p^{*})^{2} + 
       \dots \;] (p-p^{*}) \;\;\; , 
\end{eqnarray}
where $d_{o}$, $q_{o}$ and $d^{*}=D(p^{*})$, $q^{*}=Q(p^{*})$ are arbitrary
coefficients (related to material costs, wage rate, {\it etc}), $p^{*}$ is
an equilibrium price and $\delta$ is our order parameter.  
We write $D$ and $Q$ as a Taylor series
expansion with the usual linear dependence (independent of $\delta$)
plus a non-linear correction.  Higher order terms ${\cal O}(4)$ are here
neglected for small $p-p^{*}$.  In the above we might also consider
two different $\delta$s, but to reduce variables to a minimum we 
assume $D$ and $Q$ to vary similarly from linearity.  To simplify notation
we also define
\begin{equation}\label{eq:beta}
\beta_{o} \equiv q_{o}-d_{o} \;\;\; . 
\end{equation}

In the context of a simple economic model \cite{Cle84}, it is reasonable
to assume that $S_{o}$ depends linearly on the demand; {\em e.g.},
$S_{o} = \ell_{o} + \ell D$, with $\ell_{o}$ a constant and $\ell$
satisfying the constrain below.
The postulated linear dependence of the optimal stock level
$S_{o}$ on $D$ at equilibrium provides a complete economic
model as in \cite{Cle84}.
Therefore, in equilibrium (where $\frac{dp}{dt}|_{p^{*}}=0$ and
$\frac{dS}{dt}|_{S^{*}}=0$, so that demand equals supply and $S=S_{o}$),
from the above we obtain
\begin{equation}\label{eq:ps}
d^{*}-q^{*} = 0 \;\;\; ,  \;\;\; 
S^{*} = \ell_{o} + \ell ( d^{*} + d_{o} p^{*} ) \;\;\; .
\end{equation}

And after some little algebra we find that the price of one strategic
commodity is governed by the general equation
\begin{eqnarray}\label{eq:gen}
\lefteqn{ \frac{d^{2}p}{dt^{2}} +
  (\gamma \beta_{o} - \ell \lambda d_{o}) [1 - \; \frac{3\delta^{2}}{2!}
       (p-p^{*})^{2}] \frac{dp}{dt} + } \nonumber \\
 & \hspace{3.5cm} &  \lambda \beta_{o} (p-p^{*}) [
  1 - \; \frac{\delta^{2}}{2!}(p-p^{*})^{2} ] \approx 0  \;\;\; .
\end{eqnarray}
The linear case is for $\delta = 0$.  This leads to
$p(t) -p^{*} \propto A_{1} \cos (t\sqrt{\lambda \beta_{o}}) + A_{2}
sin (t\sqrt{\lambda \beta_{o}})$.

\subsection{The Simple Case $\ell \lambda d_{o} = \gamma \beta_{o}$}
 
To keep the mathematics simple we choose $p^{*}=0$ and consider $\ell$ 
to satisfy
\begin{equation}\label{eq:ell}
\ell \equiv \frac{\gamma \beta_{o}}{\lambda d_{o}} \;\;\; ,
\end{equation}
The general Eq.(\ref{eq:gen}) then reduces to
\begin{equation}\label{eq:dyn}
\frac{d^{2}p}{dt^{2}} + \lambda \beta_{o} p -
   \frac{\delta^{2} \lambda \beta_{o}}{2}p^{3}  \approx 0 \;\;\; .
\end{equation}
This is our dimensionless price adjustment equation which gives rise to
a burst as discussed later.  When $\delta \ne 0$ and
$[\lambda \beta_{o},\delta^{2} \lambda \beta_{o}/2] > 0$,
it has the well-known kink solutions
\begin{equation}\label{eq:kink}
p(t) = \pm \frac{\sqrt{2}}{\delta} \; tanh
  (\sqrt{\frac{\lambda \beta_{o}}{2}} \; t) \;\;\; .
\end{equation}
Clearly $\beta_{o}$ of Eq.(\ref{eq:beta}) must be positive.
Since in a free market economy the demand for a product (or
commodity) fall when its price increases, then it is reasonable to
assume $d_{o} < 0$ in Eq.(\ref{eq:dq}).  As the price raises, the supply
also increases; hence in general one also assumes $q_{o} > 0$.  These
conditions yield $\beta_{o} > 0$ as requested in Eq.(\ref{eq:kink})
and also $d_{o}\ell >0$.

In the case $\delta < 0$ the above function for $p(t)$ displays a sudden
decline around the equilibrium value $p^{*}$ taken to be $p(t^{*})=p(0)=0$.

\subsection{The Case $\ell \lambda d_{o} \ne \gamma \beta_{o}$}

If we consider the case in which $\ell \lambda d_{o} \ne \gamma \beta_{o}$,
the price equation of Eq.(\ref{eq:gen}) results in a Lienard-type of
equation: $p''+g_{1}(p)p'+g_{o}(p)=0$ due to the presence of the
$\frac{dp}{dt}$-term.  What we present next is a brief discussion regarding
its possible solutions.  Again, we set $p^{*}=0$.

With the aid of the substitution $\frac{dp}{dt} \equiv w(p)$, so
that $\frac{d^{2}p}{dt^{2}} \equiv w(p) \frac{dw}{dp}$, the Lienard
equation can be reducible to an Abel equation of the second kind
$ww'=f_{1}(p)w+f_{2}(p)$.  For small $\delta$, the substitution
$w(p)= p^{3}K(p)+\frac{p}{4}(\ell\lambda d_{o}-\gamma\beta_{o})$
leads to the Bernoulli equation with respect to $p=p(K)$:
\begin{equation}
 3K(p) [ \frac{\delta^{2}}{8}(\ell \lambda d_{o}-\gamma\beta_{o})
         - K(p) ] \frac{dp}{dK} = K(p) p +
    \frac{(\ell \lambda d_{o}-\gamma\beta_{o})}{4p} \;\;\; ,
\end{equation}
whose solution is
\begin{eqnarray}\label{eq:noneq}
\lefteqn{ p^{2} =
 [\frac{\delta^{2}}{8}(\ell \lambda d_{o}-\gamma\beta_{o}) -K(p)]^{-1/3}
  \; \{ \; K_{o} + \frac{(\gamma\beta_{o} - \ell \lambda d_{o})}{8}
     \; \times } \nonumber \\
 & \hspace{1.5cm} &
\times \; (\frac{-1}{3K(p)})^{2/3}
   \;\;\; _{2}F_{1}(\frac{1}{3},\frac{2}{3},\frac{5}{3},
  \frac{\delta^{2}}{8}\frac{(\ell \lambda d_{o}-\gamma\beta_{o})}{K(p)} )
  \; \} \;\;\; ,
\end{eqnarray}
with $K_{o}$ an arbitrary constant and $_{2}F_{1}$ the first, complex
hypergeometric function as arising in many physical problems.  It
converges within the unit circle
$|(\ell \lambda d_{o}-\gamma\beta_{o})/K(p)| < 2/\delta^{2}$.

By examining these equations one recognizes that if
$\ell \lambda d_{o} \ne \gamma \beta_{o}$, other behaviour might
appear for $p(t)$ (different from the one in Eq.(\ref{eq:kink})).
However, it is important to note that such possible behaviour can
essentially be found in the complex plane since the solutions of
Eq.(\ref{eq:noneq}) are driven by the $K^{-2/3}$-term and depend
whether $\gamma \beta_{o}$ is greater or smaller than $\ell \lambda d_{o}$.
A detailed analysis of such solutions is beyond the scope of this work.
We only study here the simplest dynamical economic model where all its
variables, including prices, are related to the demand and supply
functions as seen in a free market.

\section{Model and Analogous Phase Transition}

Let us now identify the behaviour of $p(t)$ with an analogous phase
transition as seen in multifractals.  Following the analogy with critical
phenomena in the time domain as proposed in \cite{Sor96}, we also consider
$t$ to be the relevant variable for the analysis of an possible existence
of an analogous critical point.  To derive a connection between our economic
model and multifractality we first briefly review multifractal phenomena
in Random Multiplicative Processes.

\subsection{Multifractality in Random Multiplicative Processes}

Multifractality emerges in random multiplicative processes for a
self-similar function $\psi$ such that is rescaled as
$\hat{\psi}(x)=e^{-L(1)x}\psi (x)$, where $L(1)$ is the generalized
Lyapunov exponent for the first moment of $\psi$ and $x$ is a space
variable for $N$-disorder fluctuations on a unit interval \cite{Pie86}.
All information about these systems is embodied in the non-zero,
positive $\psi$ measures.

To analyse the analogy of multifractality with thermodynamics one then
scales the moments $q$ of the functions $\hat{\psi}$ with respect to
segments $\l \le 1$ as
\begin{equation}
Z(q,\l) = \lim_{N \rightarrow \infty } \sum_{k=1}^{\l} \hat{\psi}^{q}_{k,N}
  \sim \l^{\tau (q)} \;\;\; ,
\end{equation}
which defines the exponents $\tau$ and $Z$ is a formal partition function.

It has been previously shown that for general random multiplicative
processes \cite{Pie86}, $\tau$ satisfies
\begin{equation}\label{eq:tau1}
\tau (q)  =  - \; (1 - q) - \; \frac{1}{h}\{ L(q) - qL(1) \} \;\;\; ,
\end{equation}
where
\begin{eqnarray}
L(q) & = &  \lim_{N \rightarrow \infty } \frac{1}{N}\ln
     \sum_{k=1}^{\l} \psi^{q}_{k,N} - h
       \;\;\; , \nonumber \\
  h   & = & \lim_{N \rightarrow \infty } \frac{1}{N}\ln \l^{-1} \;\;\; .
\end{eqnarray}

We shall use these findings to derive a connection between our economic
model variables and multifractality phenomena.  From this connection, we
shall identify all the model quantities that might exhibit multifractality
within the framework of a stochastic multiplicative process.  These
processes are known to generate power law probability density functions
\cite{Tak97,Sor98}.

\subsection{General Continuous Probability Theory}

We use next the simple continuous probability theory discussed
in \cite{Can93} that allows to explore the genesis of an "analogous" phase
transition from the point of view of a nonlinear singularity spectrum
equivalent to multifractals.  A crucial feature of this formalism is
to consider $t \sqrt{\lambda \beta_{o}/2}$ to be a {\it continuous}
random variable.  Then within the framework of general probability theory
(see, {\it e.g.},\cite{Pro69}),  
the {\it distribution function} of this random variable, defined in a line
and in terms of its {\it probability distribution} $P$,                  
can be approximated as
${\bf P} \{ \zeta_{1}< \zeta '  \leq \zeta_{2} \} =
{\cal G}(\zeta_{2} )-{\cal G}(\zeta_{1})  \approx
     \int_{\zeta_{1}}^{\zeta_{2}}  \phi (\zeta ' )
       \; d\zeta '  $,
where $\{ \}$ indicates the function interval and $\phi$ is a uniform
{\it probability density} that needs to be specified.

Inspired by well-known results for the 
probability distribution function of real economic data with
power-law tails \cite{Tak97,Sor98,Cha53,Kes73,Sorn97}, we assume
$\phi (\zeta ) \equiv (\phi_{0}/2) [1 - (\delta/\sqrt{2}) H(\zeta )]$
such that $\phi (\zeta \rightarrow + \infty )\rightarrow 0$ and
$\phi (\zeta \rightarrow -\infty )\rightarrow \phi_{0} > 0$, with $H$
given by the real solutions of a static, dimensionless Ginzburg-Landau-like
equation.   It resembles the spin-flip function of the well-known 
Glauber-Ising chain model.  Using such solutions, we shall show next that
it is possible to establish a relation with thermodynamics similarly to
multifractality phenomena.

It is at this point that we relate $H(\zeta)$ to the positive
solutions $p(t)$ of Eq.(\ref{eq:dyn}) and map
$\zeta/\zeta_{o} \rightleftharpoons t \sqrt{\lambda \beta_{o}/2}$
within the framework of general probability theory.
Following Ref.\cite{Can93}, it is then straightforward to derive an
expression for an "analogous" specific heat, $C_{\zeta}$ , for our
economic system.  To obtain such an expression we evaluate first 
the integral of {\bf P} over the range $[\zeta_{o},\zeta]$.  Thus we have
\begin{equation}\label{eq:w1}
 {\cal G}(\zeta) - {\cal G}(\zeta_{o})
    =  \frac{\phi_{0}}{2}\int^{\zeta}_{\zeta_{o}}
 \{ 1 - \frac{\delta}{\sqrt{2}} \; tanh (\frac{\zeta '}{\zeta_{o}}) \} \;
           d\zeta ' \equiv \tau (\zeta )  \;\;\; ,
\end{equation}
which, in turn, defines the function $\tau (\zeta )$.  The ${\cal G}$ 
functions satisfy ${\cal G}(\zeta ) > {\cal G}(\zeta_{o})$, or 
alternatively, $\phi (\zeta ) > \phi (\zeta_{o})$ (since
$\partial {\cal G}/\partial \zeta = \phi (\zeta )$ \cite{Pro69}),
which is in accord with the above assumption for
$\phi(\zeta \rightarrow \pm \infty)$.

Let us see next how our analysis from general probability theory would
capture multifractality.  Similarly to the dielectric breakdown model
or the Poisson growth model, where the local field is set proportional 
to the growth probability \cite{Wei93}, we assume here that the continuous 
function $\tau$ (to be identified as a free energy) is proportional to
the probability distribution ${\bf P}$ as in Eq.(\ref{eq:w1}).

After a little algebra, the above integral gives
\begin{equation}\label{eq:tau}
\tau (\zeta /\zeta_{o}) \approx (1-\zeta /\zeta_{o}) \tau (0)
        -\frac{\delta\zeta_{o}\phi_{o}}{2\sqrt{2}}
 \{ \ln cosh (\zeta /\zeta_{o}) - (\zeta /\zeta_{o}) \ln cosh (1) \}
       \;\;\; ,
\end{equation}
in which
$\tau (0)  \equiv  {\cal G}(0) -{\cal G}(\zeta_{o})
     =  - \frac{\zeta_{o}\phi_{o}}{2} \{ 1 +
          \frac{\delta}{\sqrt{2}} \Gamma_{\lambda}  \}$,
such that $\Gamma_{\lambda} \equiv -\ln cosh (1)$.
We have assumed that $\delta \Gamma_{\lambda}/\sqrt{2} << 1$, hence
$\delta < 0$.

Up to this point we have not carried out any actual numeration or scaling
of a particular fractal configuration or set as used in multifractal theory.
However, to gain further insight and explain how the present general
probability can mimics multifractal behaviour, let us now relate the
above $\tau$ to a measure within a random multiplicative process.
By mapping our results to such a measure, all other quantities that can be
scaling and are directly derived from $\tau$, such as an analogous specific
heat, will then follow.

From a comparison between Eqs.(\ref{eq:tau}) and (\ref{eq:tau1}) one can
easily identify the following terms
\begin{eqnarray}
\frac{1}{h} & \rightleftharpoons & \frac{\delta \zeta_{o} \psi_{o}}{2 \sqrt{2}}
                 \;\;\; ,  \nonumber \\
\tau (0) & \rightleftharpoons &  -1 \;\;\; ,  \nonumber \\
L(\zeta /\zeta_{o})  & \rightleftharpoons &  \ln \cosh (\zeta /\zeta_{o})
            \;\;\; .
\end{eqnarray}
It is from this simple mapping between our results and those for random
multiplicative processes that we can made an attempt to understand how the
existance of the economic model order-parameter $\delta$ (and from it,
the non-linearities in the demand and supply functions) leads to obtain
multifractal-like behaviour.

From the above mapping we deduce that if $\delta \rightarrow 0$, then
the moments $q$ of the functions $\hat{\psi}$ with respect to
segments $\l$ would vanish since $\l \rightarrow 0$.  It follows
 that a linear economic model would never exhibit
multifractal features since in this case $\tau (q) = q - 1$.
Furthermore, it is important to note that a Lyapunov exponent of the type
we derive resembles that of a random multiplicative process
with $\triangle$ described by the probability distribution
$P(\triangle ) = (1/n)\sum_{i=1}^{n} \delta(\triangle - \triangle_{i})$
\cite{Pie86,Rom88}.  By considering the actual definition of
$L(q)$, the moments of the exponential measures in a random
multiplicative process might well be related to our reduced
variable as $q \rightleftharpoons \zeta/\zeta_{o} \rightleftharpoons
t \sqrt{\lambda \beta_{o}/2}$.

\subsection{Possible Multifractal Features}

To analyse multifractal features in our economic model, when mapped
into a random multiplicative process as discussed above, we consider
standard definitions:
$\tau (\zeta /\zeta_{o}) \equiv [(\zeta /\zeta_{o}) - 1] D_{\zeta}$.
According to such definitions \cite{Lee88,Hav89,Sta90}, the
function $\tau$ represents an "analogous" free energy and $D_{\zeta}$ the
multifractal dimension.

From Eq.(\ref{eq:tau}) it follows that
\begin{equation}\label{eq:Mat1}
D_{\zeta}  \approx  -\tau (0)
       + \; \frac{\delta \zeta_{o}\phi_{o}}
       {2\sqrt{2}(1 - \zeta /\zeta_{o})}  \{
  \ln cosh (\zeta /\zeta_{o}) -(\zeta /\zeta_{o}) \ln cosh (1)  
        \} \;\;\; ,
\end{equation}
such that $\zeta \neq \zeta_{o}$.  From this relation we obtain
$D_{\zeta\rightarrow 0} = -\tau (0)$,
and $D_{\zeta \rightarrow + \infty}  =  D_{\zeta \rightarrow 0} -
\frac{\delta \zeta_{o}\phi_{o}}{2\sqrt{2}} \{ 1 +\Gamma_{\lambda} \}$;
$D_{\zeta \rightarrow -\infty}  =  D_{\zeta \rightarrow 0} +
\frac{\delta \zeta_{o}\phi_{o}}{2\sqrt{2}} \{ 1 -\Gamma_{\lambda} \}$.
If $\zeta = \zeta_{o}$, then
$D_{\zeta \rightarrow \zeta_{o}}
= D_{\zeta \rightarrow 0} - \frac{\delta \zeta_{o}\phi_{o}}{2\sqrt{2}}
\{ \Gamma_{\lambda}  + tanh (1) \}$.

Complementary to $\tau$ we also define
\begin{equation}\label{eq:cc1}
\alpha (\zeta / \zeta_{o})  \equiv
     \frac{\partial }{\partial (\zeta /\zeta_{o}) }\tau (\zeta)
 \approx D_{\zeta \rightarrow 0} - \frac{\delta \zeta_{o}\phi_{o}}{2\sqrt{2}}
     \{ \Gamma_{\lambda}  + tanh (\zeta /\zeta_{o}) \}   \;\;\; .
\end{equation}
It can be easily shown that
$\alpha_{max}\equiv \alpha (\zeta /\zeta_{o}\rightarrow-\infty)  =
D_{\zeta /\zeta_{o} \rightarrow -\infty}$, and
$\alpha_{min}\equiv \alpha (\zeta /\zeta_{o}\rightarrow+\infty)  =
D_{\zeta /\zeta_{o}\rightarrow +\infty}$.

Also according to multifractality phenomena, a possible analogy
with thermodynamics can be established by relating $\tau$ to
$f(\alpha)  \equiv  (\zeta /\zeta_{o})\alpha (\zeta /\zeta_{o}) -
\tau (\zeta /\zeta)$ via a Legendre transformation.
From this analogy, where $f$ is athe analogous "entropy",
our analytical expression for the analogous "specific heat"
of the economic system becames
\begin{equation}\label{eq:heat}
C_{\zeta}  \equiv -\; \frac{\partial^{2}\tau}{\partial
          (\zeta /\zeta_{o})^{2}}
  \approx \frac{\delta \zeta_{0}\phi_{o}}{2\sqrt{2}} \; sech^{2}
    (\zeta/\zeta_{0}) \;\;\; .
\end{equation}

\section{Discussion}

Let us see next how the type of behaviour given in Fig.1 influences the
"analogous" specific heat function $C_{q}$ defined in Eq.(\ref{eq:cqm})
and how it might characterize the onset of a crash for a real stock market
index.  In Fig.2 we plot the "analogous" specific heat $C_{q}$ of the
S\&P500 index for four different time lags $T=1,10,30,80$.  The $+++$
curve is for the 1984-1988 data, $xxx$ is for 1982-1990 and $ooo$
is for 1980-1992.

For time lags $T\rightarrow 80$, we find that the main peak of our
numerical $C_{q}$ resembles a classical (first-order) physics phase
transition at a critical point given by the main peak position.
The peak turns symmetric around the value $q = -1$.  Surprisely,
this "analogous" specific heat $C_{q}$ of the S\&P500 index also displays
a shoulder to the right of the main peak as a function of smaller time lags.
Clearly, on decreasing $T$, the presence of the shoulder is a consequence
of the large, temporal $x(t+T)-x(t)$ fluctuations in this regime.
We note that such peculiar behaviour for a double peaked specific
heat function is known to appear in the Hubbard model within the
{\it weak}-to-{\it strong} coupling regime \cite{Vol97}.

The relation in Eq.(\ref{eq:zqn}) requires $T\rightarrow \infty$ where
the shoulder tends to vanish.  It is this feature that make us believe
that a large crash for an stock market index can be characterized by
an "analogous" specific heat which resembles a classical phase
transition at a critical point as studied in multifractal physics.

We now turn to the results of our theoretical economic approach.
For $T=80$, the full line in Fig.2 represents theoretical
results for $C_{\zeta}$ from Eq.(\ref{eq:heat}) by choosing
$\zeta \rightleftharpoons q+1$ to fit the main peak position.
It can be seen that the theoretical $C_{\zeta}$ curve
resembles the phase transition features of the real S\&P500
economic data.  Our simple model for one commodity shade light
into the main observed features regarding a possible analogous
phase transition that occurs when the excess demand becomes non-linear
({\em c.f.}, cubic $p$-term in Eq.(\ref{eq:dyn})) in terms of the price
for one commodity.  We believe the width difference between both
curves, {\it i.e.} the analytical $C_{\zeta}$ and the estimated
$C(q)$ function from historical S\&P500 data, is due to the
fact that the S\&P500 price index is made of large-capitalization
stocks representing a "basket" or portfolio of commodities.

For large time lags, there is a sharp peak 
that resembles the quantitative signals measured in
multifractals \cite{Lee88,Hav89,Sta90}.  From this feature we presume the
existence of an analogous, say, critical point $\zeta^{*}$ above which
inflated prices for one strategic commodity might be found.
The maximum and minimum values of $\alpha$ in Eq.(\ref{eq:cc1}) (for more
details see also \cite{Can93}) allows for the existence of a critical point
$\zeta^{*}$ above which the infinite hierarchy of phases can be found, but
below which a single phase appears characterized by $\alpha_{max}$.
It resembles a classical phase transition at a critical point
\cite{Van98}.

Of course, the analogy between multifractality and a thermodynamic phase
transition as discussed here does not imply that the economic system has
a phase transition.  What we have shown, as a direct consequence of the
$p^{3}$-term in Eq.(\ref{eq:dyn}), is that prices can became inflated
prior to equilibrium ({\em i.e.}, $t<0$ by convention), whereas after
a sudden crash prices might devalue.  The greater $|\delta|$-values are
taken, the smaller the price reduction becomes.  If $p>0$ the amount
supplied exceeds demand and stocks accumulate whereas if $p<0$ the stocks
deplete.  Also by tuning $\delta \rightarrow 0$ in Eq.(\ref{eq:kink}), we
are able to predict that prices can decrease (or increase if $A_{2}>0$)
monotonically with $C \rightarrow 0$ for all $\zeta$.

It is a well-known fact the hill behaviour of the generalized dimensions 
$D_{q}$ for $q<0$ when using the box-counting method as in the present 
work \cite{Rie95,Rom97}.  Thus, we check if $D_{q}$ is sufficiently smooth
for $C_{q}$ to be meaningful.  In Fig.3 we represent $D_{q}$ of
the S\&P500 index for the 1980-1992 period within the range $q \ge 0$
for the time lags $T = 1, 10, 30, 80, 120$.  The total number of sequence
points analysed include 1495 points for the different $C_{q}$ curves, 1500
points for $D_{q}$ and 9864 points for the S\&P500 1980-1992 data set.  We
also fit this non-seasonal data by standard exponential smoothing
techniques, and from the fitting our error estimates for the $D_{q}$
curves is found to be less than $3\%$.  Our results for $D_{q}$ at negative
$q$ (not shown) follow a typical convergent behaviour as can be seen, for
example, in \cite{Can93}.

From Fig.3 it can be seen that, when increasing $T > 30$, $D_{q}$
is fully multifractal-like and for $T > 120$ it becomes flatter
({\it i.e.}, uniform measure) and tends to one independently of $q$,
so multifractality becomes smaller.  For lower values of $T<30$ we
find a non-monotonous decreasing behaviour of $D_{q}$, conceivable
within the double peaked form of $C_{q}$ displayed in Fig.2, which
relates to the presence of the onset of crash for the S\&P500 stock
index in Fig.1.  Using this data set, we also estimated the
multifractality strength of the time sequence by considering the limit
$1 - D_{q \rightarrow \infty}$ for the few different time lags displayed
in Fig.3.  We find that that this quantity does not follow a power-law
scaling for $T$ as in cite{Rom97} for values $T \rightarrow 1$.
This is a consequence of the complex network of trading
interactions comprised in our non-linear forms for the supply and
demand functions.

The present choice for a excess demand function of the form
$E(p) \propto (p-p^{*}) - \frac{\delta^{2}}{2}(p-p^{*})^{3}$, 
with $\delta \ne 0$ plays a key role in our description.  We have
$d^{*}=q^{*}$, hence the second price derivative
$\frac{d^{2}E}{dp^{2}} \approx 3\beta_{o}\delta^{2}p$ (or that of $D$
and $Q$) is price dependent.  It is such a behaviour of $E$ 
(independently of the sign of $\delta$),
that leads to obtain an abrupt fluctuation in the price dynamics
and characterize multifractality phenomena.

Our expressions for the demand and supply functions of Eq.(\ref{eq:dq})
are justified as follows.  As seen in Fig.4, the (commonly used) linear
$p$-dependence for $D$ and $Q$ and our
assumed non-linear form for the these functions display
similar behaviour when $|\delta p| << 1$.  Even more important, this figure
depicts the fact that as price falls, the quantity demanded for
a commodity can increase in agreement with one of the basic principles
of economy.  Since the demand curve can indeed change in a number of
ways which may not be at all obvious, similar tails to our $D$ and
$Q$ functions has also been previously hypothesized in \cite{Hen91}.
 
In real world, exceptions to the general
law of demand can take place making the $D$ curve to increase upwards
from low- to high-prices.  However these exceptions, which include
goods of conspicuous consumption -as, {\it e.g.}, certain articles
of jewelry- are not very important \cite{Thi81}.  Theoretically,
this simply would mean to set $d_{o}>0$ in our $D(p)$ function independently
of the order parameter $\delta$.  The upward dependence of $D$ can occur as
soon as there is speculation (as in precious stones mentioned but also for
market prices).  The idea is that when price increases, investors buy,
because they hope the price will keep climbing. This is called a
"trend-following" investment strategy.  On the other hand, our choice
for $Q$ (with $q_{o}>0$) also follows the typical behaviour observed in
a competitive market (where no individual producer can set his own
desired price).  That is, the higher the price, the higher the profit,
then the higher the supply.
 
It is also important to mention that the present choice for $D$ and $Q$,
and their linear $p$-dependence both lead to
the same linear relation for $D$ {\it vs.} $Q$, namely
$(D-d^{*})/d_{o} = (Q-q^{*})/q_{o}$.  Such a linear behaviour
founds frequent use in applied economics.
 
Furthermore, it is well known in econometrics that demand functions
are somewhat abstract quantities since all of these data are taken
to refer to possible events at just one moment of time \cite{Bau77}.
In particular, consumer behaviour ({\it i.e.} tastes, desires, $\dots$)
can shift a (concave up or concave down) $D$ curve (which may resemble
the tails of our $D$ and $Q$ functions).  Also, typical demand and
supply of capital look like a step function underlying forces toward
interest rates \cite{Sha90}.
 
Usually additional hypothetical
information is needed to make up realistic demand relationships
({\it e.g.}, consumer interviews).  In principle, one might also
determinate $D$ out of many data sources from different economic
sectors using standard statistical techniques for multiple regression
time series analysis.  But the effectiveness of such an approach varies
case by case.  In view of all of this, our non-linear approach for
$D$ and $Q$ may well
be placed for simulating market situations leading to a sudden decline
around equilibrium for the price of a commodity.
To this end, we add that a commodity price function
displaying an inflection point at a characteristic frequency has also
been theoretically discussed in \cite{Kro98}.
 
Another important test for our $D$ and $Q$ expressions arises
by considering possible aggregated changes in conditions of demand
(or, supply).  Because of such aggregated changes, {\it e.g.} due to
buyers' income and scale of preferences, it is always difficult
to estimate how much of the change in $D$ is due to price alone.
To know what these effects are upon $D$, and obtain the degree of
responsiveness of demand to price variations, it is necessary to
study {\it the point price elasticity of demand} curves, defined
as $\varepsilon \equiv - d \log D / d \log p$ \cite{Bau77}.
This quantity is said to be elastic (or flat) if $\varepsilon >1$ or
inelastic if $\varepsilon < 1$ depending on the demand schedule.
 
By assuming a linearly decreasing $p$-dependence for $D$ one obtains
the expression $\varepsilon = -\; \frac{1}{1+(d^{*}/d_{o}p)}$.  And if
$\delta \ne 0$ one gets
$\varepsilon \approx - \;
\frac{1-3(\delta p)^{2}/2}{[1-(\delta p)^{2}/2]+ (d^{*}/d_{o}p)}$.
Hence we find that a transition from
an inelastic to an elastic demand curve appears at
$p \approx - \; \frac{d^{*}/d_{o}}{1-(\delta p/2)^{2}}$ Whereas by
considering the common linear dependence of $D(p)$ one finds
$p \approx -d^{*}/d_{o}$.  Therefore, both approaches lead similar
results for the elasticity of demand (or supply) if $\delta \rightarrow 0$.

\section{Concluding Remarks}

We have found that within the framework of multifractal physics,
the "analogous" specific heat of the S\&P500 discrete price index displays
a shoulder to the right of the main peak for low values of time lags.
On decreasing $T$, the presence of the shoulder is a consequence
of the peaked, temporal $x(t+T)-x(t)$ fluctuations in this regime.
For large time lags ($T > 80$), we have found that $C_{q}$ displays typical
features of a classical phase transition at a critical point according to
multifractal physics.

Our simple continuos model for one commodity mimics the main observed
features of $C(q)$ for large trading time lags.  We believe the width
difference between the analytical $C_{\zeta}$ and estimated $C(q)$
curves from historical S\&P500 data is due to the fact that the S\&P500
index comprises many commodities.  From these results we conclude that
an analogous phase transition might occurs when the excess demand becomes
non-linear ({\em c.f.}, cubic $p$-term in Eq.(\ref{eq:dyn})) in terms of
the commodity price.

We have assumed that there is only one stock in which a commodity
can be stored. The market has been considered competitive so it
self-organizes to determine the behaviour of prices.  All factors determining 
$D$ and $Q$ other than $p$ are assumed to remain constant over time.
There exists a price adjustment relation that takes into account deviations
of the stock level $S$ above certain optimal level $S_{o}$
characterized by a noisy $\lambda$ parameter.
We have postulated simple non-linear forms for the quantities $D$
demanded and $Q$ supplied, with $\delta$ the order parameter,
and have neglected higher order terms in their expansion on $p-p^{*}$.
We have followed the context of other simple economic models and assumed
that the optimal stock level $S_{o}$ depends linearly on the demand 
(with $\ell$ being the slope).  We have considered $\ell$ to satisfy the
constrain in Eq.(\ref{eq:ell}), relating the economic model variables:  
$\gamma$, $\lambda$, $q_{o}$ and $d_{o}$.  We have $d^{*}=q^{*}$ 
and set the equilibrium price $p^{*}$ to 0.  We have identified the behaviour
of $p(t)$ with an analogous phase transition as seen in multifractals, by
considering $t$ to be the relevant variable and using a simple continuous
probability theory. 

We have related $t \sqrt{\lambda \beta_{o}/2}$ to a {\it continuous}
random variable and have related its probability distribution function
to our solutions for $p(t)$ given in Eq.(\ref{eq:kink}).
Our definitions for the analogous thermodynamic variables have been done
according to definitions used in multifractal physics and by mapping to
multifractality phenomena in random multiplicative processes.

Of course the scenario of a transition from, say, inflated to devalued
price changes in the time domain is pure speculation.  However a
great deal of relevant information has been extracted from the present
continuous approach which, essentially, does relay on $\delta$ only.
Our description presumes the existence of a stationary probability
function ${\bf P}$ as in Eq.(\ref{eq:w1}).  We have assumed ${\bf P}$
to be proportional to the continuous function $\tau$ (identified as a
free energy) similarly to the dielectric breakdown model or the
Poisson fractal growth models \cite{Wei93}.  We add that this type of
approximation is also used when modeling earthquakes, where the
probability function of the total number of relaxations (size) of the
earthquakes is set proportional to the energy release during an
earthquake \cite{Ola92}.

Within the framework of general probability theory (see, {\it e.g.}, 
\cite{Pro69}), a continuous arbitrary random variable can have an associated 
probability distribution (if it is discrete) or a positive probability
density (if it is continuously distributed).  The later is required to
be differentiable.  These can then be related by an integral equation of
the type used in this work to then derive the analogous thermodynamics
equations.  To do this we have proceed as follows.  We have used the 
characterization of multifractal singularities -where an analogy with
thermodynamics has been established in the literature- to derive, and
associate a meaning to our analogous quantities.  The main point to
understand is that we have not carried out any direct numeration or scaling
of particular fractal configurations or sets, but we have evaluated the
integral of the probability distribution ${\bf P}$ (for the continuous random
variable $\zeta$) assumed to be related with our $p(t)$ function and
mapped the results for $\tau$ to those of a random multiplicative process.
Our equivalent definitions for the thermodynamic variables, derived by
solving such integral, followed from the convention used in multifractal
physics in the sense that $\tau$ must not be a linear function of $\zeta$ 
(see, {\it e.g.}, \cite{Lee88}). Alternative approaches can also be found
in \cite{Nag87,Lyr98}. In all theses cases, analogous thermodynamic quantities
follow from the Legendre transform of $\tau (\zeta)$.

As discussed in \cite{Can93,Lee88}, the concept of a phase transition in
multifractal spectra was first found in the study of logistic maps,
julia sets and other simple systems.  Evidence was then found for a
phase transition in more complex random systems such as diffusion
limited aggregation.  The condition for an analogy between
multifractality and thermodynamic phase transition is that the analogous
free energy ($\alpha$ in our notation) undergoes a quite sharp jump
near a critical value $\zeta^{*}$.   For values of $\zeta < \zeta^{*}$,
the analogous free energy $\tau$ is dominated by the maximum energy term
$\alpha_{max}$ and a singular behaviour of the specific heat is found at
this point.  A well-defined analogous entropy function f($\alpha$) 
also suggests the existence of an analogous phase transition.
We have shown that the present kinetic description satisfies such 
properties obeying the constrains for $\ell$ in Eq.(\ref{eq:ell}).
This condition simplifies the analysis and corresponds to the case in which
the term for the first derivative of $p$ with respect to $t$ ({\it i.e.}, 
$\frac{dp}{dt}$) is absent in the price adjustment Eq.(\ref{eq:dyn}).

Application of multifractal analysis to discrete 1D time sequences as
derived, for example, from the cellular automaton model of a rice model
is no new (see, {\it e.g.}, \cite{Rom97}).  Nevertheless, to our best 
knowledge, we have related multifractal physics to financial time series 
for the first time.  Our main contribution has been the analysis of the
 "analogous" specific heat (or second derivatives) $C_{q}$ of the data sequence,
in conjunction with the analytical form derived from our proposed one-stock
model, which suggest typical features of a classical physics phase transition at a
critical point.  The double peaked form of $C_{q}$ is a consequence of the
presence of the onset of crash for the S\&P500 stock index.

Our work also differs from previous formalisms of multifractality of
time series in that we have analytically 
characterized multifractal singularities and its thermodynamics interpretation.
The suggested non-linear analytical forms for the supply and demand
functions of a commodity lead us to derive theoretically the observed
features of a classical phase transition
using the simplest economic model.  We believe all these
novel aspects of the topic could stimulate further investigations on this
direction and can be important to open beneficial discussions in the field
of econophysics.

\subsection*{Acknowledgments}

The author gratefully acknowledges discussions via e-mail 
with Prof. D. Sornette.

\newpage

\begin{figure}[]
\caption [A Picture]
{\protect\normalsize
Temporal fluctuations of the Standard \& Poor's 500 Index for the
period 1980-1992 as a function of trading time lags $1 \le T \le 220$.
The large, narrow fluctuations measured for $T=1$ spread out on
increasing $T$.
}
\end{figure}

\begin{figure}[]
\caption [A Picture]
{\protect\normalsize
"Analogous" specific heat $C_{q}$ of the S\&P500 index for different
time lags $T=1,10,30,80$:  The $+++$ curve is for 1984-1988 data set,
$xxx$ for 1982-1990 and $ooo$ for 1980-1992.  For $T=80$, the
full line represents theoretical results for $C_{\zeta +1}$ from
Eq.(\ref{eq:heat}).
}
\end{figure}

\begin{figure}[]
\caption [A Picture]
{\protect\normalsize
Dimension spectrum $D_{q}$ of the S\&P500 index for the 1980-1992 period
in the range $q \ge 0$ for different time lags $T = 1, 10, 30, 80, 120$.
}
\end{figure}

\begin{figure}[]
\caption [A Picture]
{\protect\normalsize
Plots of the linear $p$-dependence of $D$ and $Q$ independent of $\delta$
(dotted lines) and the assumed non-linear forms for the demand and supply
of Eq.(\ref{eq:dq}) with $|\delta p| << 1$ (full lines). 
}
\end{figure}


\newpage
 
\begin{thebibliography}{99}

\bibitem{Cha98} T. Chapman, Europhys. News {\bf 29}, 4 (1998).
\bibitem{And88} P.W. Anderson, K.J. Arrow and D. Pines editors, {\it
The Economy as an Evolving Complex System} (Addison-Wesley, 1988).
\bibitem{Con97} R. Cont, preprint cond-mat/9705075, to appear in European
Physical Journal B (1998);
\bibitem{Liu97} Y. Liu, P. Cizeau, M. Meyer, C.-K. Peng and H.E. Stanley,
Physica A {\bf 245}, 437 (1997).
\bibitem{Gha96} S. Ghashghaie, W. Breymann, J. Peinke, P. Talkner and Y.
Dodge, Nature {\bf 381}, 767 (1996).
\bibitem{Man96} R. Mantegna and H.E. Stanley, Nature {\bf 376}, 46 (1995);
{\em ibid} {\bf 383}, 588 (1996); see also Physica A {\bf 239}, 255 (1997).
\bibitem{Arn98} A. Arn\'eodo, J.-F. Muzy and D. Sornette, Eur. Phys. J. B
{\bf 2}, 277 (1998).
\bibitem{Ili97} K. Ilinsky, preprint hep-th/9710148, to appear in proceedings
of Econophysics Meeting (Budapest, 1997).
\bibitem{Sorn98} D. Sornette, Int. J. Mod. Phys. C {\bf 9}, 505 (1998). 
\bibitem{Gal97} S. Gallucio and Y.-C. Zhang, Phys. Rev. E {\bf 54}, R4516
(1996).
\bibitem{Mar98} M. Marsili, S. Maslov and Y.-C. Zhang, Physica A {\bf 253}, 
403 (1998).
\bibitem{Pot96} M. Potters, R. Cont and J.P. Bouchaud, Europhys. Lett.
{\bf 41}, 239 (1998).
\bibitem{Cal97} G. Caldarelli, M. Marsili and Y.-C. Zhang, 
Europhys. Lett. {\bf 40}, 479 (1997).
\bibitem{Sor97} D. Sornette and A. Johansen, Physica A {\bf 245}, 411 (1997).
\bibitem{Fei95} J.A. Feigenbaum and P.G.O. Freund, Int. J. Mod. Phys. B
{\bf 10}, 3737 (1996); see also Mod. Phys. Lett. B {\bf 12}, 57 (1998).
\bibitem{Sor96} D. Sornette, A. Johansen and J.P. Bouchaud, J. Phys. I
France {\bf 6}, 167 (1996).
\bibitem{Glu97} S. Gluzman and V.I. Yukalov, Mod. Phys. Lett. B {\bf 12}, 75
(1998); see also preprint cond-mat/9710336.
\bibitem{Van98} N. Vandewalle, Ph. Boveroux, A. Minguet and M. Ausloos,
Physica A {\bf 255}, 201 (1998).
\bibitem{Hal86} T.C. Halsey, M.H. Jensen, L.P. Kadanoff, I. Procaccia and
B.I. Shraiman, Phys. Rev. A {\bf 33}, 1141 (1986).
\bibitem{Mea98} P. Meakin, {\it Fractals, Scaling and Growth far from
Equilibrium} (Cambridge University Press, 1998).
\bibitem{Man99}  B. B. Mandelbrot, Sci. Am., Feb 1999, p. 50.
\bibitem{Ley96} Paper: ewp-data/9603001 - U.S. Stock Market Indices: DJIA
(1900-93) and S\&P's (1926-93). See
http://wueconb.wustl.edu/eprints/data/papers/9603/9603001.abs
\bibitem{Men87} C. Meneveau and K.R. Seenivasan, Nucl. Phys. B (Proc. Suppl.)
{\bf 2}, 49 (1987).
\bibitem{Kur97} Y. Kuramoto and H. Nakao, Phys. Rev. Lett. {\bf 78}, 4039 
(1997).
\bibitem{Ber98} A. Bershadskii, Phys. Rev. E {\bf 58}, 2660 (1998).
\bibitem{Bou98} J.P. Bouchaud and R. Cont, preprint cond-mat/9801279
(submitted to the European Physical Journal B).
\bibitem{Cle84} D.L. Clements, {\it An Introduction to Mathematical Models
in Economic Dynamics} (North Oxford Academic, Oxford, 1984).
\bibitem{Pie86} L. Pietronero and A.P. Siebesma, Phys. Rev. Lett. {\bf 57},
1098 (1986); {\it ibid} {\bf 61}, 1038 (1988).
\bibitem{Tak97} H. Takayasu, A.-H. Sato and M. Takayasu, Phys. Rev. Lett.
{\bf 79}, 966 (1997).
\bibitem{Sor98} D. Sornette, Physica A {\bf 250}, 295 (1998); see also
Phys. Rev. E {\bf 57}, 4811 (1998).
\bibitem{Can93} E. Canessa, Phys. Rev. E {\bf 47}, R5 (1993).
\bibitem{Pro69} Yu. V. Prohorov and Yu. A. Rozanov, {\it Probability
Theory: Basic Concepts, Limit Theorems, Random Processes} (Srpinger-Verlag,
Berlin, 1969).
\bibitem{Cha53} D.G. Champenowne, Economic Journal {\bf 63}, 318 (1953).
\bibitem{Kes73} H. Kesten, Acta Math {\bf 131}, 207 (1973).
\bibitem{Sorn97} D. Sornette and R. Cont, J. Phys. I France {\bf 7}, 431
(1997).
\bibitem{Wei93} W. Wei and E. Canessa, Phys. Rev. E {\bf 47}, 1243 (1993).
\bibitem{Rom88} H. E. Roman, Phys. Rev. Lett. {\bf 61}, 1037 (1988)
\bibitem{Lee88} J. Lee and H.E. Stanley, Phys. Rev. Lett. {\bf 61}, 2945
(1988).
\bibitem{Hav89} S. Havlin, B. Trus, A. Bunde and H.E. Roman, Phys. Rev.
Lett. {\bf 63}, 1189 (1989).
\bibitem{Sta90} H.E. Stanley {\it et al}., Physica A {\bf 168}, 23 (1990).
\bibitem{Vol97} D. Vollhardt, Phys. Rev. Lett. {\bf 78}, 1307 (1997).
\bibitem{Rie95} R. Riedi, J. Math. Ana. Appl. {\bf 189}, 462 (1995). 
\bibitem{Rom97} R. Pastor-Satorras, Phys. Rev. E {\bf 56}, 5284 (1997).
\bibitem{Hen91} J.V. Henderson and W. Poole, {\it Principles of
Economics}, D.C. Heath and Company, International Student Editions (1991).
\bibitem{Thi81} G.L. Thirkettle, {\it Basic Economics}, M\&E Handbooks,
4th Edition, The Chaucer Press, UK, (1981).
\bibitem{Bau77} W.J. Baumol, {\it Economic Theory and Operations
Analysis}, 4th Edition, Prentice-Hall Inc., London (1977).
\bibitem{Sha90} W.F. Sharpe and G.J. Alexander, {\it Investments},
4th Edition, Prentice Hall, New Jersey, USA (1990).
\bibitem{Kro98} A. Krouglov, EconWPA preprints, ewp-mac/9802023 and references
therein.
\bibitem{Ola92} Z. Olami, H.J.S. Feder and K. Christensen, Phys. Rev. Lett.
{\bf 68}, 1244 (1992).
\bibitem{Nag87} T. Nagatani, J. Phys. A {\bf 20}, L381 (1987).
\bibitem{Lyr98} M.L. Lyra and C. Tsallis, Phys. Rev. Lett. {\bf 80}, 53 (1988).
\end{thebibliography}
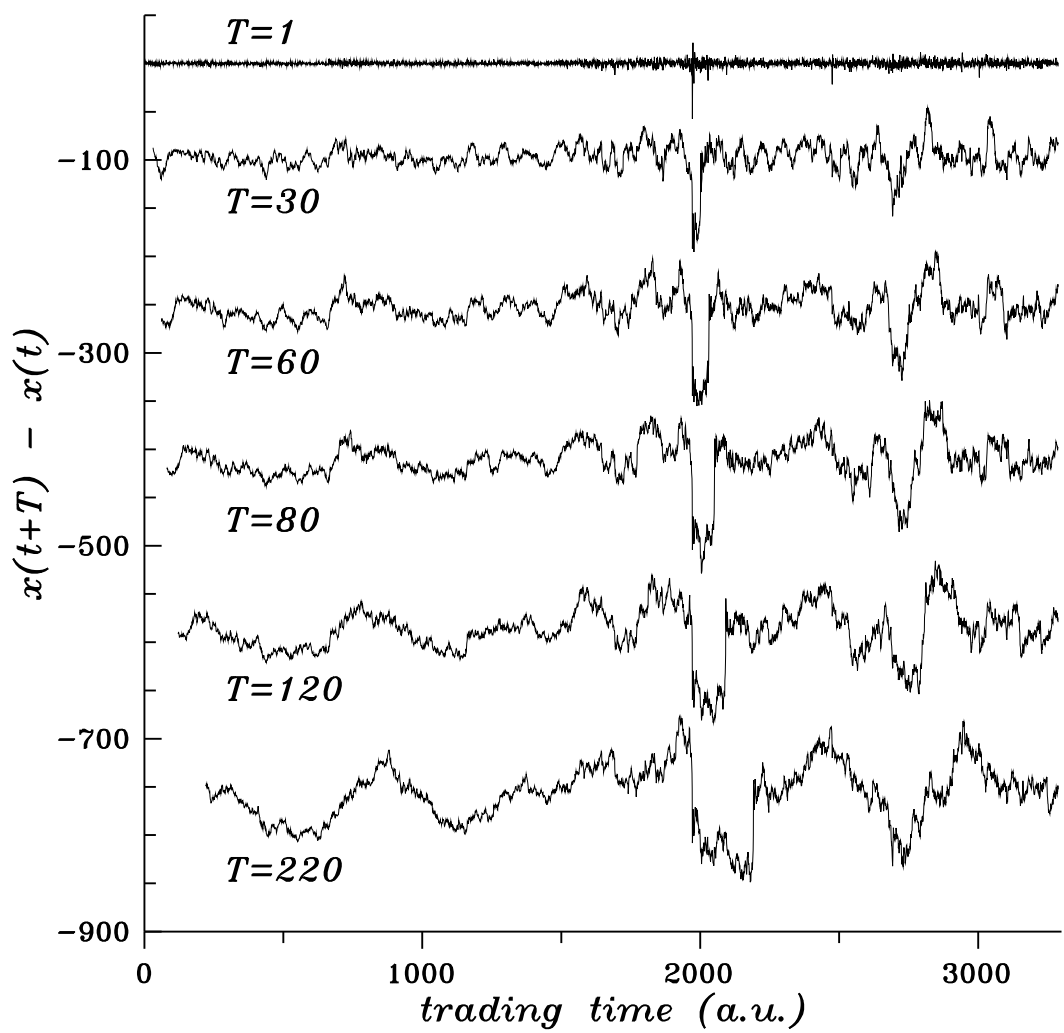
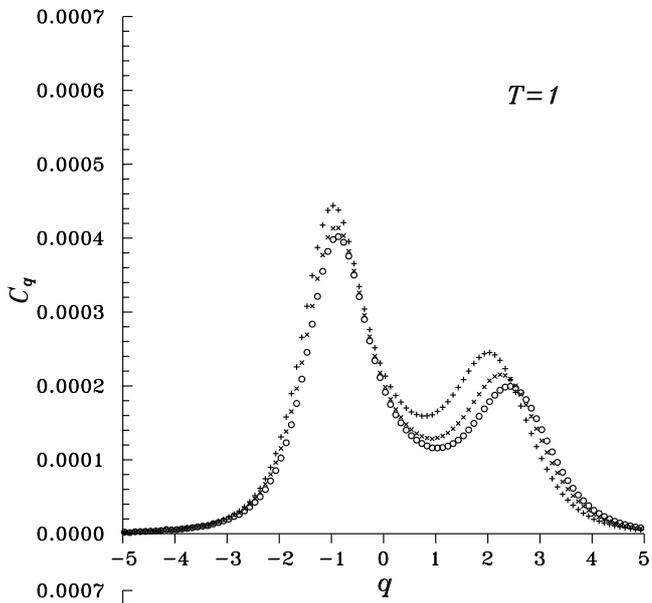
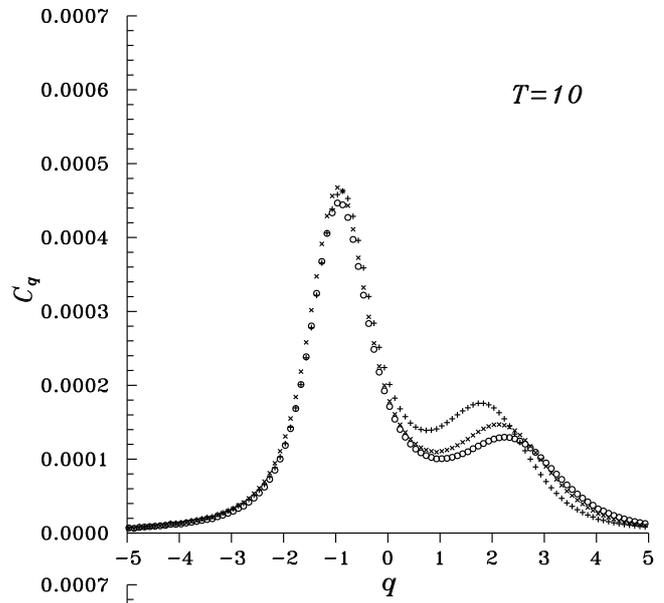
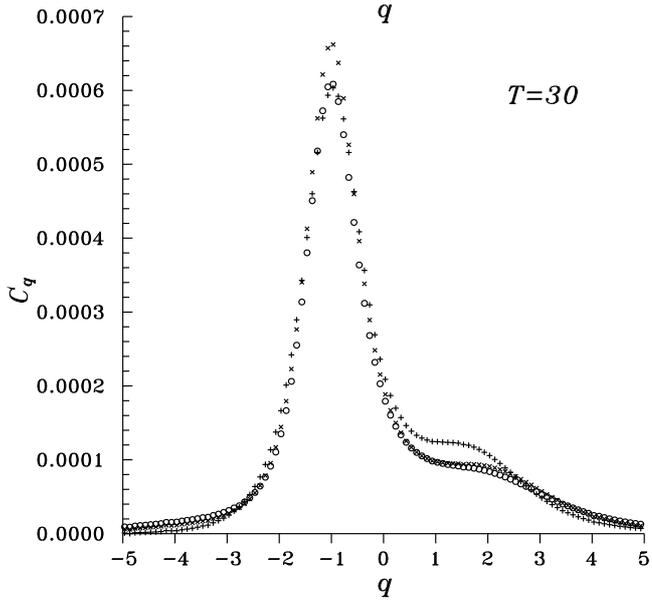
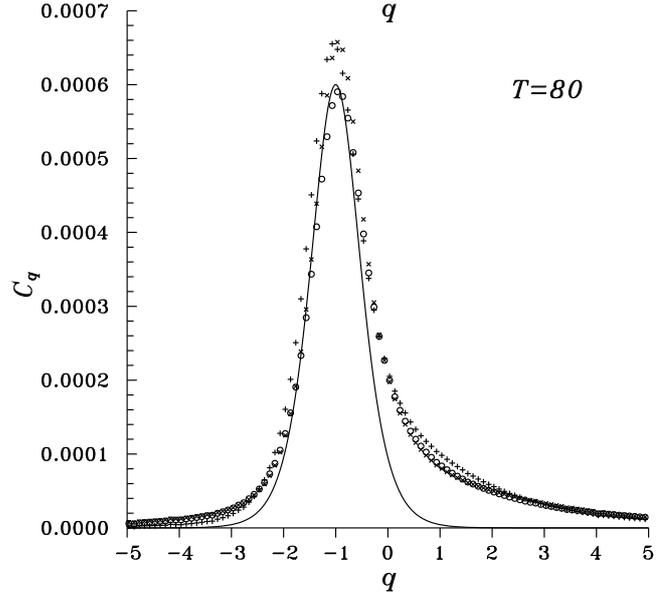
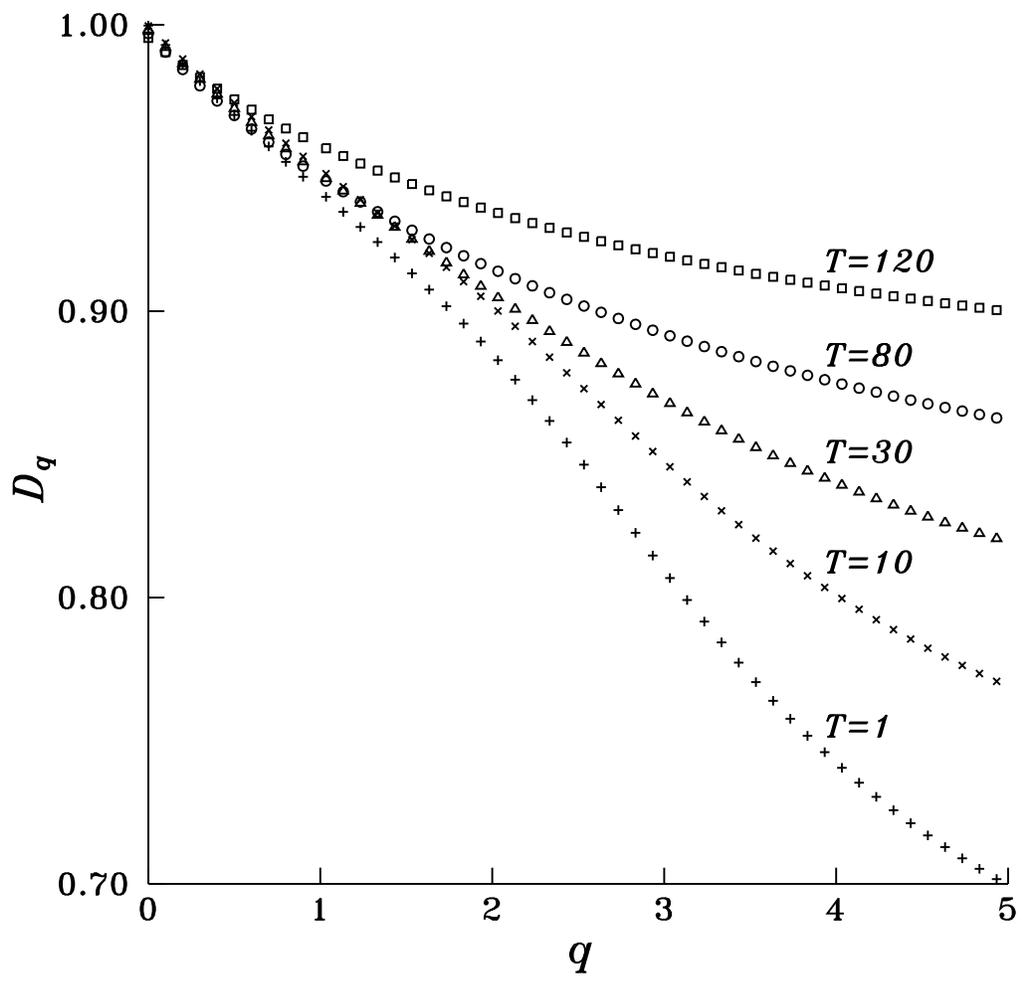
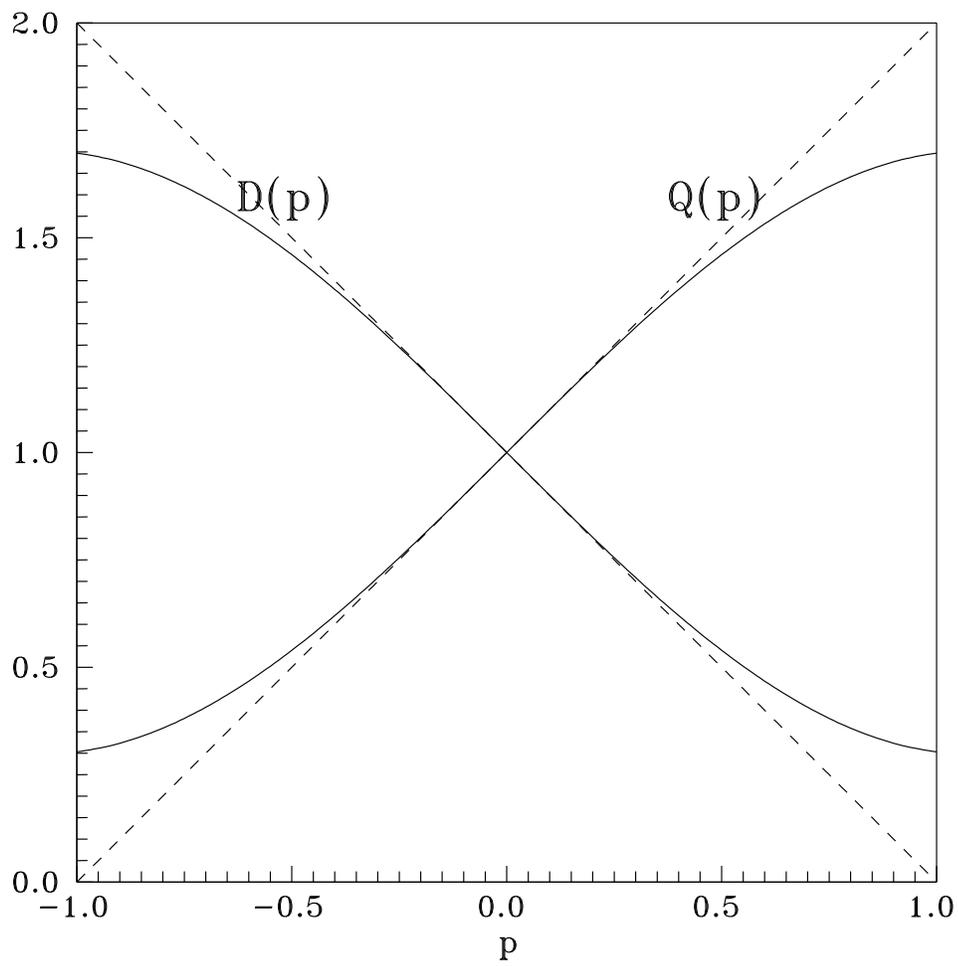
\end{document}